\documentstyle[11pt,aaspp4]{article}

\received{1997 April 16}
\accepted{1997 August 14}

\begin{document}     

\newcommand{\squig}{$\sim$}
\newcommand{\squigleq}{\mbox{$^{<}\mskip-10.5mu_\sim$}}
\newcommand{\squiggeq}{\mbox{$^{>}\mskip-10.5mu_\sim$}}
\newcommand{\decsec}[2]{$#1\mbox{$''\mskip-7.6mu.\,$}#2$}
\newcommand{\decsectim}[2]{$#1\mbox{$^{\rm s}\mskip-6.3mu.\,$}#2$}
\newcommand{\decmin}[2]{$#1\mbox{$'\mskip-5.6mu.$}#2$}
\newcommand{\asecbyasec}[2]{#1$''\times$#2$''$}
\newcommand{\aminbyamin}[2]{#1$'\times$#2$'$}

\title{The Probable Optical Counterpart of the Luminous X-ray Source in
NGC\,6441\footnote{\ Based on observations with the NASA/ESA Hubble
Space Telescope, obtained at the Space Telescope Science Institute,
which is operated by the Association of Universities for Research in
Astronomy, Inc., under NASA contract NAS5-26555.}
}
\author{Eric W. Deutsch, Scott F. Anderson, and Bruce Margon}
\affil{Department of Astronomy, 
       University of Washington, Box 351580,
       Seattle, WA 98195-1580\\
       deutsch@astro.washington.edu; anderson@astro.washington.edu;
       margon@astro.washington.edu}

\author{and\\ \vskip .1in Ronald A. Downes}
\affil{Space Telescope Science Institute,
       3700 San Martin Drive,
       Baltimore, MD 21218\\
       downes@stsci.edu}

\begin{center}
Accepted for publication in The Astrophysical Journal\\
To appear in volume 493, 1998 February 1\\
{\it received 1997 April 16; accepted 1997 August 14}
\end{center}

\begin{abstract}

We report results from {\it Hubble Space Telescope} WFPC2
imaging of the field of the luminous, bursting  X-ray source in the globular
cluster NGC\,6441.  Although the X-ray position is known to a precision
of a few arcseconds, this source is only $\sim6''$ from the cluster
center, and the field contains hundreds of stars within the 3$''$ X-ray
error circle, making it difficult to isolate the optical counterpart.
Nevertheless, our multicolor images reveal a single, markedly UV-excess
object with $m_{336}=19.0,\ m_{439}=19.3$, within the X-ray error
circle.  Correcting for substantial reddening and bandpass differences,
we infer $B_0=18.1$,\ $(U-B)_0=-1.0$, clearly an unusual star for a
globular cluster.

Furthermore, we observe an ultraviolet intensity variation of 30\% for
this object over 0.5 hr, as well as an even greater
variation in $m_{439}$ between two {\it HST} observations taken
approximately one year apart.  The combination of considerable
UV-excess and significant variability strongly favors this object as
the optical counterpart to the low-mass X-ray binary X\,1746--370.

With a group of five optical counterparts to high-luminosity globular
cluster X-ray sources now known, we present a homogeneous set of {\it
HST} photometry on these objects, and compare their optical properties
with those of field low-mass X-ray binaries.  The mean $(U-B)_0$ color
of the cluster sources is identical to that of the field sources, and
the mean $M_{B_0}$ is similar to bursters in the field.  However, the
ratio of optical to X-ray flux of cluster sources seems to show a
significantly larger dispersion than that of the field sources.

\end{abstract}

\keywords{globular clusters: individual (NGC 6441) --- stars: neutron ---
ultraviolet: stars --- X-rays: bursts --- X-rays: stars}

\clearpage
\section{INTRODUCTION}

For over two decades it has been known that globular cluster X-ray
sources are overabundant with respect to those in the field (Katz 1975;
Clark 1975).  In fact, nearly $10^{-1}$ of the luminous X-ray sources
(L$_{\rm X}\, \squiggeq\, 10^{36}$ erg s$^{-1}$) in the Galaxy are found in
clusters.  The question of whether these cluster sources are somehow
different in nature from sources in the field, or they are the same but
the globular cluster environment enhances their formation probability,
remains unanswered.

Thus far only four optical counterparts to globular cluster luminous
X-ray sources are known: those in M\,15, NGC\,6712, and NGC\,6624 seem
secure, and a good candidate which still requires confirmation has recently
been provided in NGC\,1851 (Deutsch et al. 1996a).  The difficulty in
identifying optical counterparts is almost certainly primarily due to
the extreme crowding in the clusters, which limits the utility of
ground-based identification programs.  All but the optically bright
star AC\,211 in M\,15 have required the resolution of {\it HST} for
confident identification.  We have obtained multicolor {\it Hubble
Space Telescope} images of several globular clusters containing X-ray
sources, with the aim of resolving optical counterparts in crowded
fields and studying their spectra, and report here on the results from
WFPC2 images in NGC\,6441.  This paper extends the preliminary results
and original counterpart identification suggested in Deutsch et al. (1996b).

The intense X-ray source in the core of NGC\,6441,
X\,1746--370 (= 4U\,1746--37), is a ``classical",
high X-ray luminosity (L$_{\rm X}\sim 6\times 10^{36}$ erg~s$^{-1}$)
burster observed for many years, by many satellites.
However, it is distinguished by being one of the few cluster sources
with a periodic X-ray modulation, exhibiting a 5.7~hr period of
$\sim5\%$ amplitude (Parmar et al. 1989, Sansom et al.  1993).  The
source is located just $6''$ $(\sim 1\ r_c)$ from the cluster center,
and thus the optical field is very crowded. To our knowledge there has
not even been a past suggestion of a candidate optical counterpart
before this work, despite some effort (Bailyn et al. 1988).  The small
number of high luminosity cluster sources for which details are
available already presents a confusing picture:  despite similar
X-ray luminosities, their optical luminosities and orbital periods
differ by several orders of magnitude.  The orbital period of this
object is intermediate between the NGC\,6624/6712 sources {\it vs.}
M\,15 AC\,211.

\section{OBSERVATIONS AND DATA REDUCTION}

\subsection{Planetary Camera Imagery}

On 1994 August 8, we obtained {\it HST} WFPC2 images centered on
NGC\,6441 through the F336W (50 s, 2~$\times$~500~s) and F439W (50 s,
500 s) filters, similar to Johnson U
and B, respectively.  These frames are free of any artifacts or
scattered light from the nearby bright star HR 6630 ($V\sim3$) which
often hinders observations of this cluster.  The images have been
processed through the standard data reduction pipeline at ST\,ScI.
Further reduction was performed with software written in IDL by E.W.D.
or available in the IDL Astronomy User's Library (Landsman 1993).

The sets of exposures were combined with a cosmic-ray rejection
algorithm.  We use DoPHOT (Schechter et al. 1993) to photometer stars
in all four chips of the WFPC2.  The PSF-fit magnitudes
are calibrated using aperture photometry of isolated stars, where aperture
corrections are taken from Table 2(a) in Holtzman et al. (1995b).  The
photometric measurements have not been corrected for geometric
distortions, nor is any correction for charge transfer efficiency
losses (Holtzman et al. 1995b) applied, as there is sufficient
charge on the chips to minimize this effect.  We use the photometric
zero points for the STMAG system from Table 9 ($Z_{STMAG}$) in Holtzman
et al.  (1995a).  Systematic errors for all magnitudes due to
uncertainties in detector performance and absolute calibration are
\squig 5\%.  As discussed later, severe crowding may lead to larger
overall photometric uncertainties for some objects.

\subsection{Astrometry}

In order to determine the position of the X-ray source coordinates on
the {\it HST} PC image, we establish a coordinate system based on the {\it
HST} Guide Star Catalog (GSC) reference frame.  We begin with the
astrometric solution from a digitized Quick V image used to generate
the {\it HST} GSC (Lasker et al. 1990); this frame,
obtained directly from ST\,ScI, contains the astrometric solution in
the image header.  By centroiding 38 stars in the Quick V image and the
corresponding objects in a ground-based CCD image, kindly
provided by G. Jacoby, the astrometric reference frame is transferred
to the CCD image with an error in the solution of \decsec{0}{03}.

There are only three stars which are sufficiently isolated to be
well-resolved in the ground-based data within the PC field of view, so
instead of transferring the solution from the CCD to the PC image, we
correct the astrometric information originally provided in the PC
header for the average observed offset in position for these three
stars (coordinates measured in the PC frame are adjusted by
$\Delta \alpha=+$\decsec{0}{2} and $\Delta
\delta=+$\decsec{0}{5}).  The final result is that coordinates may be
determined on the PC image to within \squig \decsec{0}{1} in the GSC
reference frame.  However, there may well be some systematic offset,
$\sigma \sim$ \decsec{0}{5}, from frames based on other reference
catalogs (Russell et al. 1990).  It should be noted that NGC\,6441 is
near the edge of the Quick V plate, where systematic errors can be
higher.

Figure 1 shows the entire \asecbyasec{33}{33} F336W PC frame, with our
position for the 3$''$ radius X-ray error circle overlaid, as well as the
\asecbyasec{8}{8} region around the circle.  The X-ray coordinates are
derived from {\it Einstein} HRI observations by Grindlay et al. (1984),
who report the 90\%\ confidence error as 3$''$.

\section{DISCUSSION}

\subsection{Color-Magnitude Diagram}

From all four WFPC2 chips we select \squig 17,000 stars for which we
have obtained good photometry, and plot their colors and magnitudes in
Fig. 2.  The scatter in the diagram is largely due to the extreme
crowding, even by {\it HST} standards, in this cluster.  There is also
doubtless some contamination from field stars for this low Galactic
latitude cluster.

We label five objects on the diagram.  Star U1 is the most
UV-excess object within the X-ray error circle (but not in the
cluster!).  U7, a star of similar color and magnitude, but well outside
the error circle, is discussed below.  Stars B1 and B2 are bluer than
the other 245 objects in the error circle (except U1), but seem to be
part of a large population of stars which compose a formidable blue
tail of the horizontal branch; they are not likely to be related to the
X-ray source.  We also indicate a rare object, the central star of the
globular cluster planetary nebula JaFu 2 (Jacoby \& Fullton 1994;
Jacoby et al. 1996, 1997).  These five objects are discussed further below.

An isochrone from Bertelli et al. (1994) is added to the diagram for
comparison.  We select a 14 Gyr isochrone with [Fe/H$]=-0.40$, the
closest available to the published metallicity for NGC\,6441,
[Fe/H$]=-0.53$, and apply distance modulus $(m-M)_0=15.15$ (Djorgovski
1993).  The isochrone $U$ and $B$ magnitudes are converted to $m_{336}$
and $m_{439}$, where the corrections depend on color; these corrections
are estimated using the STSDAS {\it synphot} package.  Finally, we
apply reddening to the isochrone, although we find that E$(B-V)=0.50$
yields a better fit to the giant branch than the published value
E$(B-V)=0.42$.  However, the uncertainties in the metallicity and
filter conversion may easily account for this small shift.
Qualitatively, the turnoff and giant branch are modeled well by the
isochrone.  The known planetary nebula central star lies near the
predicted track.  The $M_{B_0}$ scale uses a filter correction
$B-m_{439}=0.65$.  This correction changes slightly with stellar color;
0.65 is appropriate for F type through the hottest stars, while 0.4 is
more appropriate for M0 stars.

Typically a metal-rich cluster such as NGC\,6441 has a very red
horizontal branch, and so it is surprising  that our diagram exhibits a
bimodal horizontal branch distribution with such an extensive blue
tail.  These peculiarities in NGC\,6441 were recently discussed in
results from a study of several metal-rich globular clusters with {\it
HST} by Rich et al. (1997), where some possible explanations are
explored.

NGC\,6441 also appears to contain a sizable population of
supra-horizontal branch stars, which can be seen in our diagram as well
as in Rich et al.  (1997).  The bluest supra-horizontal branch stars
and members of the blue tail may be the principal contributors to the
significant but unresolved far-ultraviolet radiation detected with {\it
IUE} by Rich et al.  (1993).  It is possible that some of these objects
are not cluster members, but they do seem to follow the same radial
distribution as other cluster stars.

Finally we point out a group of stars even bluer than the extreme blue
horizontal branch stars.  Two noteworthy members of this population are
U1 and JaFu 2, our candidate for the optical counterpart to the X-ray
source and a planetary nebula central star, respectively.  It remains
to be seen if the nature of the other stars in this group is equally
exotic.

\subsection{Star U1}

The UV-excess object which we denote U1 is the only unusual star in
the X-ray error circle, and it is therefore an obvious optical
counterpart candidate to the globular cluster bursting X-ray source
X\,1746--370.  Figure 3 shows a \asecbyasec{3}{3} section of both the
F336W ($U$) and F439 ($B$) images centered on this object.  Its
$M_{B_0}=3.0\pm0.1$ and $(U-B)_0=-1.0\pm0.1$ may be compared with the
typical X-ray burster in the field which has
$1\,\squigleq\,M_B\,\squigleq\,5$ and ($U-B)=-1.0$ (van Paradijs
1995).  Therefore, not only is the color unusual for a globular cluster
star, but the color and magnitude naturally associate the object with
known X-ray burster optical counterparts.  Based on our astrometric
solution derived from the {\it HST} GSC, we find coordinates for Star
U1 $\alpha(2000)={\rm17^h50^m}$\decsectim{12}{61},
$\delta(2000)={\rm-37^\circ03'}$\decsec{06}{5}.  The measurement
uncertainty here is negligible compared to the external uncertainties
discussed in \S 2.2.

To explore possible variability of Star U1, we perform aperture photometry
on this object and two other nearby objects of similar magnitude and
color for all publicly-available {\it HST} WFPC2 and (pre-COSTAR) FOC
observations of this field, and present the results in Table 1.
Magnitudes are in the STMAG system for both instruments (therefore
permitting intercomparison) and $1\sigma$ errors are provided.  The
considerable reddening correction has not been applied to these
magnitudes.

We find clear evidence for \squig 30\% variability between our two long
F336W exposures, taken only 27 min apart.  Another short F336W exposure
and an FOC F342W image also indicate large variability, although the
photometric errors are higher and filter/instrument differences could
be responsible for at least part of the apparent discrepancy.
Additional evidence for the variability of Star U1 comes from a \squig
60\% change in $m_{439}$ between our observations and similar
observations in the archive taken 1.1 yr later for another program.
During the same period, measurements for Stars U7 and B2 (UV and blue
stars respectively of similar magnitude) are consistent with no
variability within the uncertainties.  Such large-amplitude variability
for U1 clearly adds considerable confidence that it is indeed the
optical counterpart to the low-mass X-ray binary (LMXB).

We note that the X-ray source in NGC\,6441 shows variability on
timescales of 0.25~hr at amplitudes of \squig 15\% (Parmar et al.
1989).  Assuming our suggested identification is correct, our data
indicate ${\rm L_X/L_{opt}\sim 10^3}$, implying that X-ray fluctuations
may be promptly reprocessed into detectable optical variability.  Thus
the similarity of the observed X-ray and optical variability timescales
may not be coincidental.

Despite the unusual color and variability, this identification is not
completely secure, however, due to a large number of other UV-excess
objects in the cluster, as can be seen in the color-magnitude
diagram.  In order to estimate the probability that one of these
UV-excess stars has fallen in the X-ray error circle by chance, we note
that there are 21 objects with $(m_{336}-m_{436})<0.0$ in our sample
of \squig 17,000 stars.  Therefore, approximately $10^{-3}$ of the
stars with $m_{439}<22$ in the cluster have marked UV excess, and as we
measure 245 stars within the error circle, we estimate a \squig 30\%
probability that one of the UV-excess population objects would fall by
chance in the error circle and might be mistaken for the optical
counterpart to the X-ray source.  The hazards of these {\it a
posteriori} probability estimates are well-known, so this result must
surely be regarded as qualitative rather than quantitative, but the
conclusion can also be reached visually from the distribution of UV
sources seen in Fig. 4.  We find the fraction of ultraviolet-excess
stars in the cluster intriguing; it exceeds by $\sim5\times$ an
analogous observation for NGC\,1851 by Deutsch et al. (1996a).

Previous workers have noted the marked dispersion in optical
luminosities of globular cluster X-ray source counterparts.  If U1 is
indeed the correct counterpart of the NGC\,6441 source, it falls at an
optical luminosity similar to that of the NGC\,6624 source and
intermediate between the sources in NGC\,6712 (Anderson et al. 1993)
and NGC\,1851 (Deutsch et al. 1996a), which are remarkably faint, and that
of M\,15 AC\,211.

We also attempted {\it HST} spectroscopic observations of U1 with the
Faint Object Spectrograph on 1995 July 25, using G160L and G570H
gratings, which cover the $1150-2500$ \AA\ and $4600-6800$
\AA\ regions, respectively.  The resulting G570H spectrum exhibits only spectral
features typical of a late G star.  The absolute flux level in the
spectrum is about 4$\times$ higher than $m_{555}$ for Star U1 in Table
1, but is consistent with the total F555W light in a \decsec{0}{5}
diameter circle (the size of the FOS aperture used) centered on this
object as measured on a WFPC2 image.  We conclude that most of the
light in this spectrum is most probably contributed by nearby sources
other than Star U1.

The count rate in the G160L spectrum is so low that flux coming from
the source cannot be distinguished from an imperfect scattered light
subtraction.  We set upper limits of order $3\times10^{-17}$ erg cm$^{-2}$
s$^{-1}$ \AA$^{-1}$ at 1600 \AA\ and $1\times10^{-17}$ erg cm$^{-2}$
s$^{-1}$ \AA$^{-1}$ at 2100 \AA.  These upper limits are $\sim3\times$
lower than the measurements obtained from FOC and WFPC2 observations of
this source.  This may indicate very large amplitude variability, for
which of course there is already direct imaging evidence.  However, the FOS
observations were executed by making a blind offset from a nearby
bright star, and it cannot be guaranteed that Star U1 was actually in
the aperture for either the G160L or G570H spectra.

\subsection{Planetary Nebula JaFu 2}

Planetary nebulae are rare in globular clusters; only 4
have been discovered despite considerable effort (Jacoby \& Fullton
1994; Jacoby et al. 1997).  We measure a magnitude and color for the
central star of the planetary nebula JaFu 2 in NGC\,6441 (Jacoby \&
Fullton 1994; Jacoby et al. 1996) which are remarkably similar (Fig. 2)
to that of Star U1, our proposed X-ray source optical counterpart.
This similarity naturally leads to the possibility that Star U1 may be
of comparable nature to JaFu 2, but we believe that the similarity must
be coincidental.  The planetary nebula star lies quite close
to predicted evolutionary tracks, and so its observed properties are
consistent with photospheric radiation from a hot, single star. The
X-ray source, on the other hand, must be a low-mass binary system
containing a neutron star, as X-ray bursts are observed (Li \& Clark
1977; Sztajno et al. 1987), and thus its integrated light contains
contributions from the compact star, the accretion disk, and the
companion.  The observed magnitude and colors of Star U1 are consistent
with those of bursters in the field.  Therefore, despite the unusual
nature of both objects, one could have predicted in advance that they
could both occupy the same place in the cluster color-magnitude diagram
and yet be physically dissimilar.

The natures of two of the UV-excess objects in this cluster have now
probably been determined, but the other UV-excess sources remain
unexplained.  Since they are not luminous in X-rays, they are not
similar to Star U1.  They may well be planetary nebula central stars,
where the surrounding nebula has long since dispersed.  However, the
cases of Star U1 and JaFu 2 have shown that a similarity in color and
magnitude in this diagram does not allow us to infer similarity in
nature.

\section{THE LUMINOUS GLOBULAR CLUSTER X-RAY SOURCES}

Of the 12 luminous LMXBs associated with globular clusters listed in
van Paradijs (1995), five now have confirmed or likely optical
counterparts (see Deutsch et al. 1996a for references).  In order to
make a comparison between this sample of globular cluster sources and
sources in the field, we have extracted images of each cluster from the
{\it HST} data archive and measured $m_{336}$ and $m_{439}$ for each
globular cluster (GC) LMXB optical counterpart; these values are
presented in Table 2.

Photometric errors are typically dominated by the \squig5\%\ systematic
calibration uncertainties.  The uncertainty for Star A in NGC\,1851 is
higher, 0.1 mag, due to a presumably unrelated companion \decsec{0}{12}
distant (Deutsch et al. 1996a).  Observations of the source in
NGC\,6624 are complicated by an even closer blend (separation \decsec{0}{08},
King et al. 1993), completely unresolvable with WFPC2, although easily
resolved in an F430W FOC image.  We estimate $m_{439}$ for both stars
in WFPC2 observations by measuring the total light in the blend and
using the flux ratio of the two stars in the F430W FOC image.  We then
estimate $m_{336}$ for the X-ray source counterpart by subtracting the
amount of light expected from the companion, assuming that the latter
has the same color as other stars with its $m_{439}$.  The uncertainty
in the measurements for this optical counterpart is \squig0.2 mag,
comparable to the observed amplitude of variability (Anderson et al.
1997).  Indeed, it should be noted that we have obtained only
instantaneous magnitudes for all these objects, and the sources are known
to be variable in optical or ultraviolet light, except in the case of
NGC\,1851 (for which there has not yet been an adequate light curve
study).  Nonetheless, the results are entirely sufficient for our
purpose here.

\textheight 9.30in

With this homogeneous set of photometric measurements from WFPC2
observations, E$(B-V)$ and $\rm{(m-M)_0}$ values from Djorgovski (1993),
and approximate color-dependent conversion relations between STMAG
system magnitudes and Johnson $U$,$B$ (derived using the STSDAS {\it
synphot} package), we calculate $B_0$, $(U-B)_0$, and $M_{B_{0}}$,
and list them in columns 7-9 of Table 2.  In column 10 we list
$\xi=B_0+2.5{\rm\,log\,F_X(\mu Jy})$, the parameter used by van
Paradijs \& McClintock (1995; hereafter vPM) to characterize the ratio
of X-ray to optical flux.  We use ${\rm F_X}$ values from van Paradijs
(1995).

By virtue of their cluster membership, the distance to and foreground
reddening of these GC LMXBs can be determined easily, unlike those in
the field.  This allows a comparison of $(U-B)_0$, $M_{B_{0}}$, and $\xi$
for GC sources with those of field LMXBs for which these quantities are
known.  vPM find for the field sources an average
$(U-B)_0=-0.97\pm0.17$ ($1\sigma$).  All five GC LMXBs measured in
Table 2 fall within $1\sigma$ of this average.

vPM find an average $(B-V)_0=-0.09 \pm 0.14$ for the field LMXB, so we
assume $M_B=M_V$ in the comparison of absolute magnitudes, as we do not
have $V$ data available for several objects.  This is likely to be a
good assumption, since the $(U-B)_0$ colors show a small dispersion and
for Star U1, $(B-V)_0=0.0$ (based on the 1995 September 12 data in
Table 1).  In addition, $(B-V)_0\sim0.15$ for AC\,211 (Ilovaisky et al.
1993), the reddest object in our table.  $M_V$ for bursting sources in
the field ranges considerably in vPM, from 1 to 5 with a slight
overabundance near $M_V=1$.  Our distribution in $M_B$ for the 5 GC
LMXBs, which are all bursting sources, is similar: a range from 1 to 6,
but with no overabundance near $M_V=1$.  In fact, there is evidence
that the globular clusters sources are on average less luminous: while
70\%\ of field bursters have $M_V<2.5$, only one of globular cluster
bursters has $M_B<2.5$.  However, with only seven and five sources in
each group, respectively, the difference may not be
significant.

vPM noted that the bursters they considered (mainly in the field) as a
group appear to have systematically lower optical luminosity
counterparts than other LMXBs.  The data on additional counterparts to
the globular cluster bursters discussed here add further evidence for
such a difference.

Finally we examine how field and globular cluster LMXBs compare in
$\xi$.  The source in M\,15 has long been known to be optically
overluminous and is already singled out at the bright end of vPM's
histogram.  The other four GC sources span the entire main distribution
of field LMXBs; they do not seem to show a peak near $\xi=22$ as do the
field sources, although of course the sample is small.

In summary, the sample of GC LMXBs optical counterparts obtained thus
far appears to be identical in $(U-B)_0$ and similar in $M_B$ to the
sample of field bursting sources for which comparison is possible.  The
uniformity in $(U-B)_0$ might well be expected, as the light at these
wavelengths may be dominated by a hot accretion disk, whose spectral
slope is rather insensitive to temperature in this regime.  However,
the GC sources show a very broad distribution in $\xi$ rather than the
peak of $\xi=21.8\pm1.0$ ($1\sigma$) seen by vPM; in fact, only one
of the five cluster sources falls within $1\sigma$ of the
average of the field LMXBs.

Gnedin \& Ostriker (1997) have recently reevaluated destruction rates
of globular clusters, and concluded that these timescales are more
rapid than previously inferred. A substantial fraction of all initial
globular clusters may have been destroyed in a Hubble time, and thus a
large fraction of the current stellar population of the bulge may come
from clusters. LMXBs are the most luminous stars in globular clusters,
and might be useful as a tracer of this hypothesis (see also Grindlay
1985). Indeed, Ostriker (1997) points out that luminous (near-Eddington
limited) LMXBs are overabundant in the bulge relative to other galactic
locations, just as they are overrepresented in globular clusters. The
similarity that we infer here between the optical properties of LMXBs
in clusters and those in the bulge is compatible with this interesting
concept for the origin of the bulge sources, although clearly the
sample size is small. As the bulge LMXBs obviously cannot have been
radiating at their current X-ray luminosities for a Hubble time, there
are also missing evolutionary parts of this picture.

\section{CONCLUSION}

We have examined WFPC2 images in NGC\,6441 for the optical counterpart
to X\,1746--370.  A color-magnitude diagram of \squig 17,000 stars in all
4 chips reveals a bimodal horizontal branch with a long blue tail.
Nearly two dozen strongly UV-excess objects are also seen in the CM diagram.  Our
astrometry, based on the {\it HST} GSC, places the 3$''$ X-ray error
circle accurately onto the PC frame.  Only one of the UV-excess objects
or otherwise unusual stars falls within this error circle.  The
$M_{B_0}=3.0$, $(U-B)_0=-1.0$ for Star U1 is typical of the average
properties of bursting X-ray sources in the field.  We detect a \squig 30\%\ variability
between our F336W images separated by 0.5 hr, and a \squig
60\%\ variability between our F439W frames and F439W images taken 1.1
yr later.  The color and variability combine to make U1 a very strong
candidate for the optical counterpart.

We do note that there is a sizable population of UV-excess stars in the
cluster and calculate a \squig 30\%\ probability that a member of this
population would have fallen in the X-ray error circle by chance.  An
FOS spectrum ostensibly of Star U1 detects no significant UV flux with an upper limit
$3\times$ lower than fluxes obtained from imaging, indicating either
very high UV variability or target acquisition failure.  We plan {\it
HST} STIS time-resolved spectroscopy to confirm the candidate and lead
to better understanding the nature of this object.

With five cluster optical counterparts now known, we present a
homogeneous set of {\it HST} $m_{336}$ and $m_{439}$ measurements, and
calculate $M_{B_{0}}$, $(U-B)_0$, and $\xi$ for each object.  A
comparison with sources in the field for which these values are
determined show remarkably good agreement in $(U-B)_0$, a similarity in
$M_{B_{0}}$ for bursters in the field and in globular clusters (but
with burster counterparts perhaps less luminous than other field
LMXBs), and apparently larger dispersion in $\xi$ for the globular
cluster X-ray sources.  One-quarter century after their discovery in
X-rays, enough optical counterparts are finally known (largely thanks
to {\it HST}) that an initial study of the ensemble optical properties
of this interesting class of object is now possible.

\acknowledgments

We thank George Jacoby and collaborators for providing ground-based CCD
images of NGC\,6441 and for sharing information about JaFu 2.  Support
for this work was provided by NASA Grant NAG5-1630.


\clearpage

\begin{figure}
\plotone{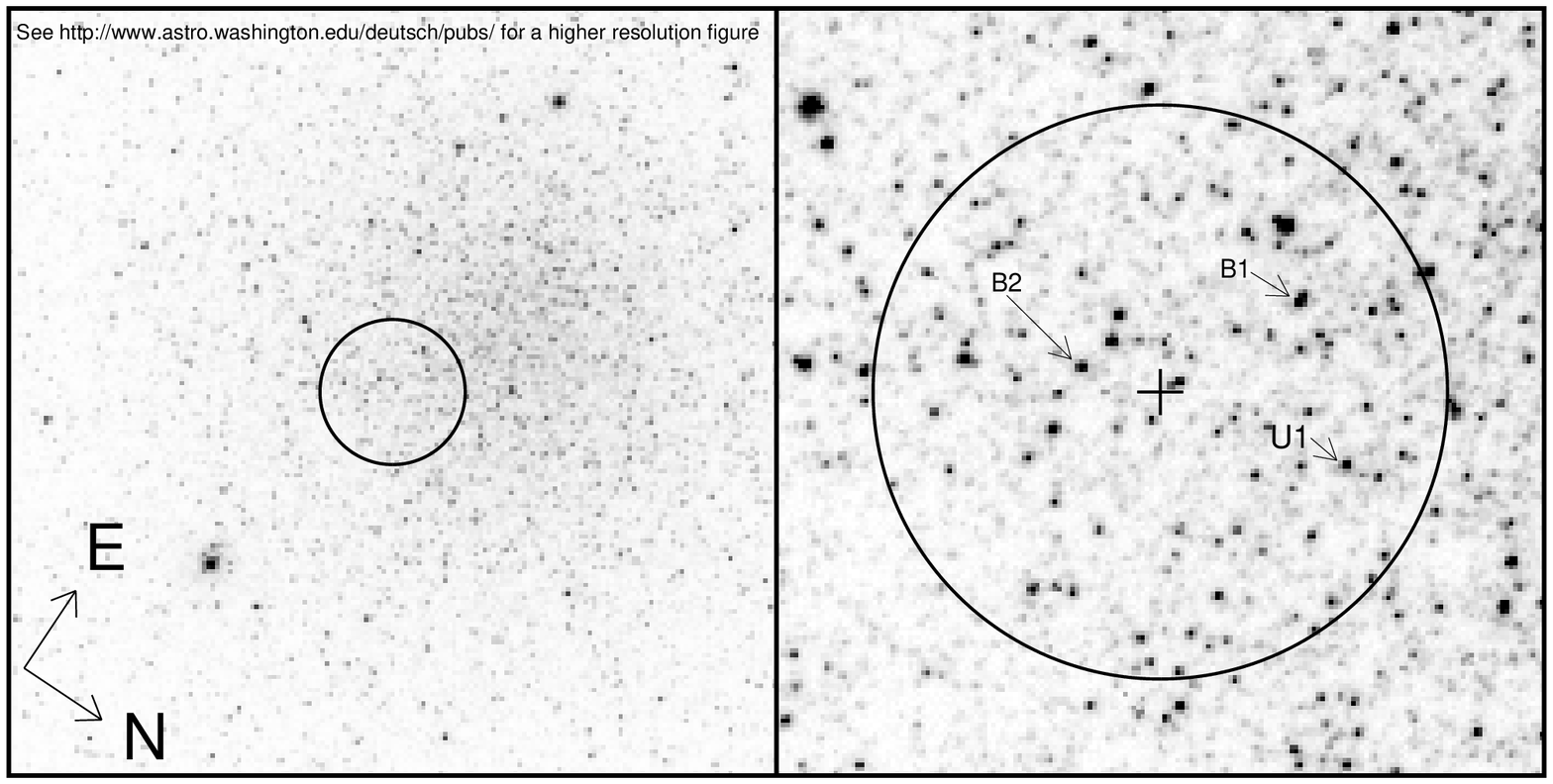}
\caption{{\it Left panel}: \asecbyasec{33}{33} {\it HST} WFPC2 PC image
of the core of NGC\,6441 with the F336W ($U$) filter.  A 3$''$ radius error
circle has been drawn about the {\it Einstein} HRI coordinates for the
luminous bursting X-ray source X\,1746--370.  {\it Right panel}:
\asecbyasec{8}{8} section of the same image, centered at our optical
position for the X-ray coordinates (indicated with a cross).
Ultraviolet-excess object U1, which we offer as a likely
optical-counterpart candidate, is indicated along with two blue stars
(labeled B1 and B2) which also fall within the error circle.}
\end{figure}

\begin{figure}
\figurenum{3}
\plotone{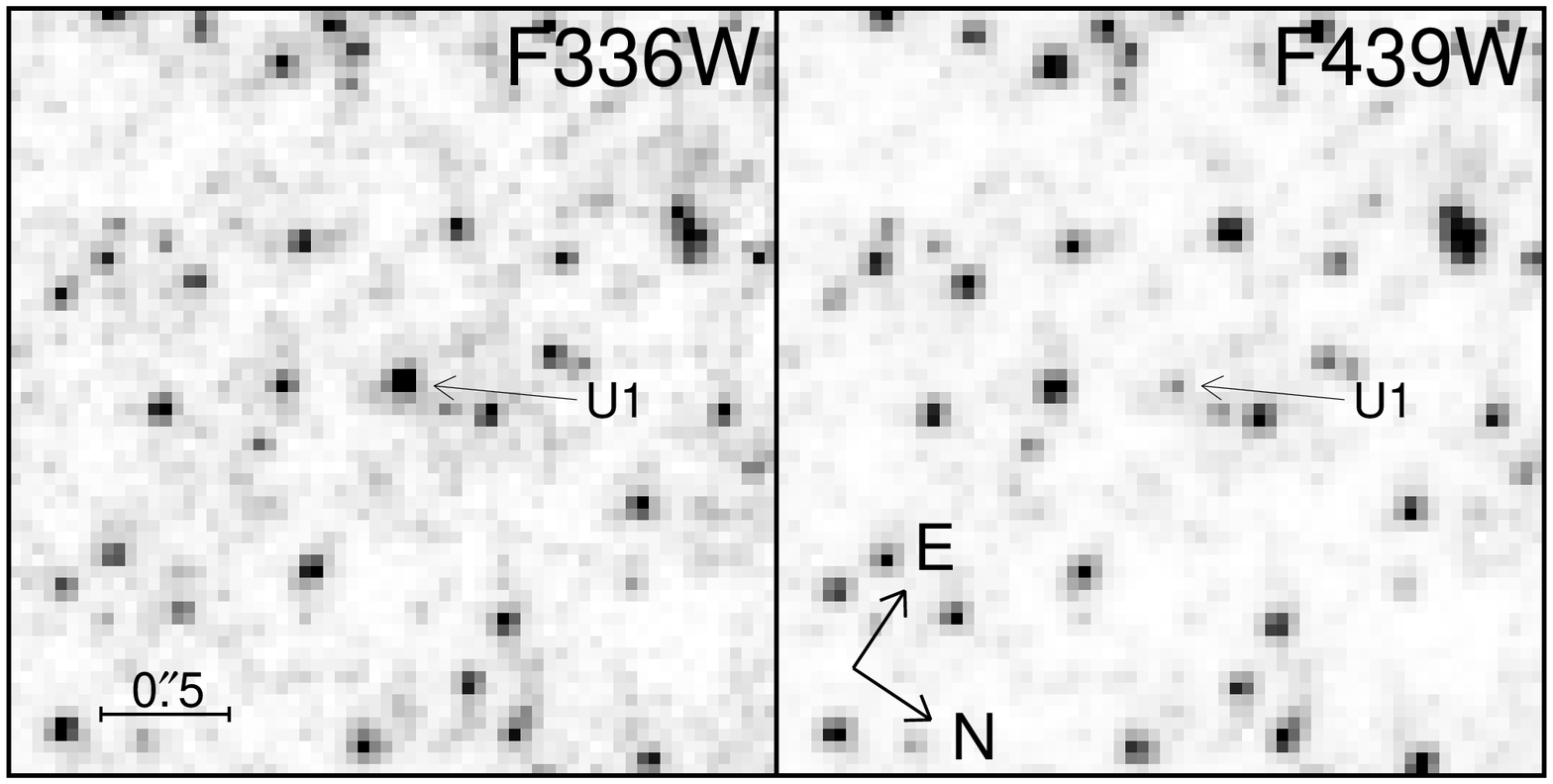}
\caption{\asecbyasec{3}{3} sections of the F336W ($U$) and F439W ($B$)
images centered on Star U1, which we select as a candidate optical
counterpart due to significant UV excess and variability.  These images
were obtained on 1994 August 8 (cf. Table 1).}
\end{figure}

\begin{figure}
\figurenum{2}
\plotone{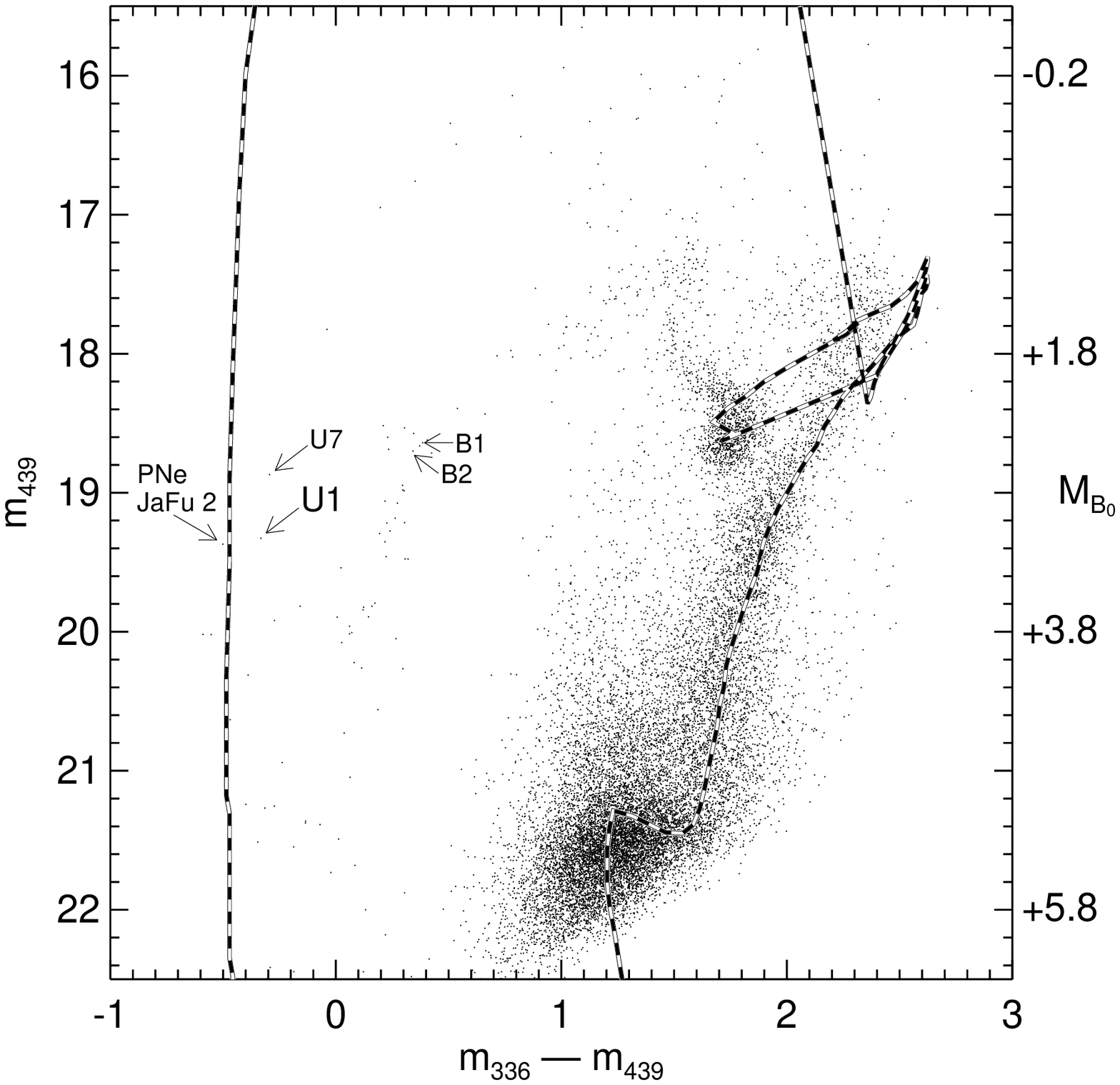}
\caption{Color-magnitude diagram for \squig 17,000 stars on all 4 chips
of the WFPC2 exposures.  Magnitudes are in the STMAG system; see text
for details.  Objects which are discussed in the text are labeled:
Star U1, which we select as a candidate optical counterpart due to
significant UV excess and variability; the central star of cluster-member
planetary nebula JaFu 2; blue Stars B1 and B2, which also fall within
the X-ray error circle but are probably not related to the X-ray
source; Star U7, one of several objects with similar color and
magnitude to U1.  Note the surprisingly well-populated branch of
extreme blue horizontal branch stars typified by B1 and B2.  An
isochrone from Bertelli et al. (1994) is also overlaid; see text
for more details.}
\end{figure}

\begin{figure}
\figurenum{4}
\plotone{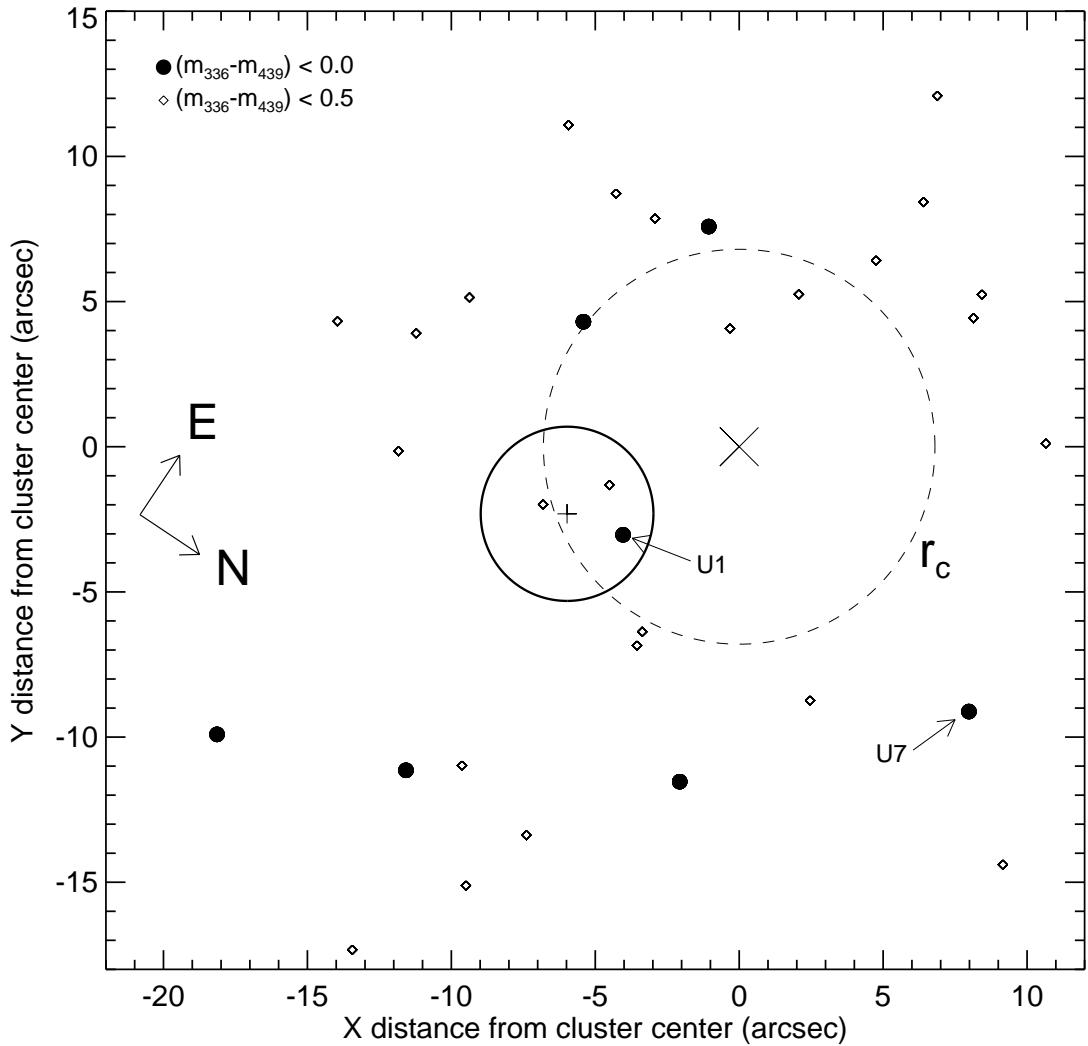}
\caption{The locations of the UV stars $(m_{336}-m_{436})<0.0$ and blue
stars $(m_{336}-m_{436})<0.5$ on the PC image (compare to Fig. 1).  The
{\it Einstein} X-ray error position (cross) and 3$''$ radius error
circle are shown (solid line).  The cluster center and core radius are
shown with a $\times$ and dashed-line circle, respectively.  Stars U1,
B1, and B2 are within the error circle, but a considerable population
of both UV and blue stars can be seen throughout the entire PC image.
Note that the bluest objects do not appear to have a marked central
concentration.}
\end{figure}

\clearpage

\oddsidemargin -0.7in
\textwidth 7.3in
{\small

\begin{deluxetable}{lccccccccc}
\tablenum{1}
\tablecolumns{10}
\tablewidth{541.39305pt}
\tablecaption{Photometry for selected objects in current observations and archival {\it HST} data of NGC\,6441}
\tablehead{
\colhead{Filter (Camera)} &
\colhead{Obs Date} &
\colhead{Start Time} &
\colhead{Exp. Time} &
\multicolumn{2}{c}{Star U1} &
\multicolumn{2}{c}{Star U7} &
\multicolumn{2}{c}{Star B2} \\
\multicolumn{3}{c}{} &
\colhead{(s)} &
\colhead{(mag)} & \colhead{(err)} &
\colhead{(mag)} & \colhead{(err)} &
\colhead{(mag)} & \colhead{(err)}
}
\startdata
F140W (FOC)     & 93/09/28 & 03:29  &\,\ 996  & 19.20 & 0.24 &  19.03 & 0.15 & $>20.8$  & 0.8  \nl

F210M (FOC)     & 93/09/28 & 04:51  &\,\ 996  & 20.22 & 0.36 &  19.56 & 0.18 &  20.01 & 0.25 \nl

F218W (WFPC2)   & 95/09/12 & 06:13  &\,\ 500  & 19.73 & 0.18 &  19.16 & 0.15 &  20.19 & 0.26 \nl
                &          & 06:24  &   1000  & 19.94 & 0.15 &  19.44 & 0.11 &  20.14 & 0.15 \nl

F336W (WFPC2)   & 94/08/08 & 13:37  &\ \ \ 50 & 19.12 & 0.07 &  18.61 & 0.05 &  19.09 & 0.07 \nl
                &          & 13:40  &\,\ 500  & 18.87 & 0.02 &  18.62 & 0.01 &  19.07 & 0.02 \nl
                &          & 14:07  &\,\ 500  & 19.19 & 0.03 &  18.65 & 0.02 &  19.08 & 0.02 \nl

F342W (FOC)     & 93/09/28 & 05:13  &\,\ 496  & 18.81 & 0.09 &  18.79 & 0.04 &  19.07 & 0.10 \nl

F439W (WFPC2)   & 94/08/08 & 13:52  &\ \ \ 50 & 19.19 & 0.10 &  18.91 & 0.05 &  18.67 & 0.05 \nl
                &          & 13:55  &\,\ 500  & 19.28 & 0.04 &  18.87 & 0.02 &  18.65 & 0.03 \nl
                & 95/09/12 & 06:01  &\ \ \ 50 & 18.77 & 0.06 &  18.91 & 0.06 &  18.57 & 0.06 \nl
                &          & 06:04  &\,\ 160  & 18.80 & 0.04 &  18.78 & 0.04 &  18.61 & 0.04 \nl
                &          & 06:08  &\,\ 160  & 18.75 & 0.04 &  18.80 & 0.04 &  18.63 & 0.04 \nl

F555W (WFPC2)   & 95/09/12 & 05:55  &\ \ \ 14 & 18.94 & 0.06 &  19.25 & 0.06 &  18.95 & 0.06 \nl
                &          & 05:57  &\ \ \ 50 & 18.95 & 0.04 &  19.26 & 0.06 &  18.94 & 0.07 \nl
\enddata
\end{deluxetable}


{\small

\begin{deluxetable}{lccccccccc}
\tablenum{2}
\tablecolumns{10}
\tablewidth{578.57048pt}
\tablecaption{Photometry for globular cluster LMXB optical counterparts from {\it HST} data}
\tablehead{
\colhead{Optical Counterpart} &
\colhead{$(m-M)_0$\tablenotemark{a}} &
\colhead{E$(B-V)$\tablenotemark{a}} &
\colhead{[Fe/H]\tablenotemark{a}} &
\colhead{$m_{439}$\tablenotemark{b}} &
\colhead{$(m_{336}-m_{439})$\tablenotemark{b}} &
\colhead{$B_0$} &
\colhead{$(U-B)_0$} &
\colhead{$M_{B_0}$} &
\colhead{$\xi$\tablenotemark{c}} 
}
\startdata
   NGC 1851 Star A & 15.43 &  0.02 & -1.29 & 20.46 & -0.73 & 21.03 & -0.98 &  5.60 & 22.97 \nl
  NGC 6441 Star U1 & 15.15 &  0.42 & -0.53 & 19.30 & -0.30 & 18.18 & -1.06 &  3.03 & 21.94 \nl
NGC 6624 King Star & 14.54 &  0.28 & -0.37 & 17.99 & -0.37 & 17.46 & -0.98 &  2.92 & 23.46 \nl
   NGC 6712 Star S & 14.16 &  0.46 & -1.01 & 19.88 & -0.19 & 18.58 & -1.02 &  4.42 & 20.70 \nl
        M15 AC 211 & 15.11 &  0.05 & -2.17 & 15.49 & -0.45 & 15.91 & -0.81 &  0.80 & 17.86 \nl
\enddata
\tablenotetext{a}{\,Cluster properties from Djorgovski (1993) and Peterson (1993)}
\tablenotetext{b}{\,Photometric uncertainties are discussed in the
text, but most of these objects are known to be variable and the rest
are likely to be as well.}
\tablenotetext{c}{\,Optical photometry from this work; ${\rm F_X}$ from van Paradijs (1995)}
\end{deluxetable}

}


\clearpage
\begin{thebibliography}{}

\bibitem[]{} Anderson, S. F., Margon, B., Deutsch, E. W., \& Downes,
R. A. 1993, \aj, 106, 1049

\bibitem[]{} Anderson, S. F., Margon, B., Deutsch, E. W., Downes, R. A.,
\& Allen, R. G. 1997, \apj, 482, L69

\bibitem[]{} Bailyn, C. D., Grindlay, J. E., Cohn, H., \& Lugger, P. M.
1988, \apj, 331, 303

\bibitem[]{} Bertelli, G., Bressan, A., Chiosi, C., Fagotto, F. \& Nasi, E.
1994, \aaps, 106, 275

\bibitem[]{} Clark, G. W. 1975, ApJ, 199, L143

\bibitem[]{} Deutsch, E. W., Anderson, S. F., Margon, B., \& Downes, R. A.
1996a, \apj, 472, L97

\bibitem[]{} ------------. 1996b, BAAS, 28, 1328

\bibitem[]{} Djorgovski, S. 1993, in ASP Conf. Ser. 50, Structure
and Dynamics of Globular Clusters, ed. S. G. Djorgovski \& G. Meylan
(San Francisco:  ASP), 373

\bibitem[]{} Gnedin, O. Y., \& Ostriker, J. P. 1997, ApJ, 474, 223

\bibitem[]{} Grindlay, J. E. 1985, in IAU Symp. 113, Dynamics of
Star Clusters, ed. J. Goodman \& P. Hut (Dordrecht: Kluwer), 43

\bibitem[]{} Grindlay, J. E., Hertz, P., Steiner, J. E., Murray, S. S.,
\& Lightman, A. P. 1984, \apj, 282, L13

\bibitem[]{} Holtzman, J. A., Burrows, C. J., Casertano, S., Hester, J. J., 
Trauger, J. T., Watson, A. M., \& Worthey, G. 1995a, PASP, 107, 1065

\bibitem[]{} Holtzman, J. A., et al. 1995b, PASP, 107, 156

\bibitem[]{} Ilovaisky, S. A., Auri\`ere, M., Koch-Miramond, L.,
Chevalier, C., Cordoni, J.-P., \& Crowe, R. A. 1993, A\&A, 270, 139

\bibitem[]{} Jacoby, G. H., \& Fullton, L. 1994, BAAS, 26, 1384

\bibitem[]{} Jacoby, G. H., Morse, J., Fullton, L., Kwitter, K.,
\& Henry, R. B. C. 1996, BAAS, 29, 1353

\bibitem[]{} Jacoby, G. H., Morse, J., Fullton, L. K., Kwitter, K. B.,
\& Henry, R. B. C. 1997, in preparation

\bibitem[]{} Katz, J. I. 1975, Nature, 253, 698

\bibitem[]{} King, I. R. et al. 1993, ApJ, 413, L117

\bibitem[Landsman 1993]{lan93} Landsman, W. B. 1993, in ASP Conf. Ser.
52, Astronomical Data Analysis Software and Systems II, ed. R. J.
Hanisch, R. J. V.  Bissenden, \& J. Barnes (San Francisco: ASP), 256

\bibitem[Lasker et al.,\ 1990]{las90} Lasker, B. M., Sturch, C. R.,
McLean, B. J., Russell, J. L., Jenkner, H., \& Shara, M. M. 1990, \aj,
99, 2019

\bibitem[]{} Li, F., \& Clark, G. 1977, IAUC 3095

\bibitem[]{} Ostriker, J. P. 1997, paper delivered to Ostriker Festschrift, Princeton NJ, May 1997

\bibitem[]{} Parmar, A. M., Stella, L., \& Giommi, P. 1989, A\&A, 222, 96

\bibitem[]{} Peterson, C. J. 1993, in ASP Conf. Ser. 50, Structure
and Dynamics of Globular Clusters, ed. S. G. Djorgovski \& G. Meylan
(San Francisco:  ASP), 337

\bibitem[]{} Rich, R., M., et al. 1997, ApJ, 484, L25

\bibitem[]{} Rich, R. M., Minniti, D., \& Liebert, J. 1993, ApJ, 406, 489

\bibitem[Russell et al.,\ 1990]{rus90} Russell, J. L., Lasker, B. M.,
McLean, B. J., Sturch, C. R., \& Jenkner H. 1990, \aj, 99, 2059

\bibitem[]{} Sansom, A. E., Dotani, T., Asai, K. \& Lehto, H. J. 1993, MNRAS, 262, 429

\bibitem[]{} Schechter, P., Mateo, M., \& Saha, A. 1993, PASP, 105, 1342

\bibitem[]{} Sztajno, M., Fujimoto, M. Y., van Paradijs, J., Vacca, W. D.,
Lewin, W. H. G., Penninx, W., \& Tr\"{u}mper, J. 1987, MNRAS, 226, 39

\bibitem[]{} van Paradijs, J. 1995, in X-Ray Binaries, ed. W. H. G.
Lewin, J. van Paradijs, \& E. P. J. van den Heuvel (Cambridge:
Cambridge Univ.), 536

\bibitem[]{} van Paradijs, J., \& McClintock, J. E. 1995, in X-Ray
Binaries, ed. W. H. G.  Lewin, J. van Paradijs, \& E. P. J. van den
Heuvel (Cambridge: Cambridge Univ.), 58 (vPM)

\end{thebibliography}
\end{document}